\documentclass[conference, twocolumn]{IEEEtran}

\usepackage{graphics,graphicx} % for pdf, bitmapped graphics files
\usepackage{epsfig} % for postscript graphics files

\usepackage{amsmath}
\usepackage{amssymb}
\usepackage{tikz}
\usepackage{amsthm}
\usepackage{comment}
\usepackage{mathcomSTEP}
\usepackage{subcaption}
\usepackage{setspace}
\usepackage[utf8]{inputenc}
\usepackage{cite}

\usepackage{amsmath}
\usepackage{steinmetz}
\usepackage{color}
\usepackage{graphicx} % for pdf, bitmapped graphics files
\usepackage{epsfig} % for postscript graphics files

\usepackage{amsmath}
\usepackage{amssymb}
\usepackage{tikz}
\usepackage{amsthm}
\usepackage{comment}

\newcommand{\Y}{{\bf Y}}

\newcommand{\I}{I}
\newcommand{\mi}[2]{{\I}\left(#1 \, ; #2 \right)}
\newcommand{\micnd}[3]{{\I}\left(\left. #1 ; #2 \,\right| #3\right)}

 \newcommand{\h}{\mathsf h}

 \newcommand{\entcnd}[2]{{\h}\left(\left. #1 \,\right| #2\right)}

\newcommand{\Exp}{\mathsf E}
\newcommand{\expect}[1]{{\Exp}\left[#1\right]}
\newcommand{\expcnd}[2]{{\Exp}\left[\left. #1 \,\right|\, #2\right]}

\newcommand{\Var}{\mathsf{Var}}
\newcommand{\variance}[1]{{\Var}\left[#1\right]}

\newcommand{\markov}{\mathrel\multimap\joinrel\mathrel-%
	\mspace{-9mu}\joinrel\mathrel-}

\newcommand{\bpcu}{\rm bpcu}

\allowdisplaybreaks

\title{
Capacity Outer Bound and Degrees of Freedom of  Wiener Phase Noise Channels with Oversampling
}

\author{
	\IEEEauthorblockN{Luca Barletta}
	\IEEEauthorblockA{
		Politecnico di Milano, Italy\\
		\texttt{luca.barletta@polimi.it}
	}
	\and
	\IEEEauthorblockN{Stefano Rini}
	\IEEEauthorblockA{
		National Chiao Tung University, Taiwan\\
		\texttt{stefano@nctu.edu.tw}
	}
}

\begin{document}
\maketitle
\begin{abstract}
The discrete-time Wiener phase noise channel with
an integrate-and-dump multi-sample receiver is studied.
A novel outer bound on the capacity with an average input power constraint
is derived as a function of the oversampling factor.
This outer bound yields the degrees of freedom for the scenario in which the oversampling factor grows with the transmit power $P$ as $P^{\al}$.
The result shows, perhaps surprisingly, that the largest pre-log that can be attained with phase modulation at high signal-to-noise ratio is at most $1/4$.
\end{abstract}

\section{Introduction}

In the discrete-time Wiener phase noise (WPN) channel, the channel input is affected by both
 additive white Gaussian noise (AWGN) and multiplicative WPN.
The Wiener phase process can be used to model a number of random phenomena: from imperfections in the oscillator
circuits at the transceivers, to slow fading effects in wireless environments or laser imperfections in optical communications.
For the WPN channel, the sampled output of the filter
matched to the transmit filter does not always represent a
sufficient statistic~\cite{barletta2014continuous}, and oversampling does help in achieving
higher rates over the continuous-time channel~\cite{martalo2013information,ghozlan2013wiener}.
For this reason, it is of interest to study the effect of oversampling on the maximum achievable rates~\cite{ghozlan2014phase}.
In this paper, we study the discrete-time Wiener phase noise with oversampling channel, a model obtained by sampling the output of the continuous-time phase noise channel faster than the symbol frequency.
For this model, we determine a novel outer bound on capacity and provide the generalized degrees of freedom (GDoF) for the scenario in which the oversampling factor grows
to infinity as $P^{\al}$ where $P$ is the transmit power.
This result shows that, even in the high signal-to-noise ratio (SNR) regime, it is not possible to attain more than a pre-log factor of 1/4 through phase modulation.

%\newpage
\subsubsection*{State of the Art}
The literature on channel affected by both additive noise and phase noise considers three models: (i) the continuous-time model, (ii) the discrete-time model and (iii) the
discrete-time model with oversampling.
%

%continuous
For the \emph{continuous-time} case,  the joint effect of phase noise and additive white Gaussian noise is first
considered in~\cite{foschini1988characterizing}.
In~\cite{goebel2011calculation}, the authors investigate
white (Gaussian) phase noise for which they observe a “spectral
loss” phenomenon induced by white phase noise.
The continuous-time channel in the presence of white
noise is proposed and discussed in~\cite{barletta2014continuous}.
Here it is shown that, for linear modulation, the output of the baud-sampled filter
matched to the shaping waveform represents a sufficient statistic.
Bounds on the SNR penalty for the case of Wiener phase
noise affecting the channel input are developed in~\cite{barletta2014signal}.

%discrete
The \emph{discrete-time} phase noise channel  is obtained by considering a discrete-time WPN process sampled at symbol frequency.

This model was first studied in~\cite{lapidoth2002phase}: here the high SNR capacity is derived using duality arguments.
The authors of~\cite{barletta2012JLT} propose a numerical method of precise evaluation of information rate bounds for this model.
In \cite{khanzadi2015capacity} the authors derive closed-form approximations to capacity which are shown to be tight through numerical evaluations.

%oversampling
Finally, in the \emph{discrete-time model with oversampling} multiple samples for every input symbol are obtained in output.
This model was first considered in \cite{Ghozlan2014ISIT}
where it is shown that, if the number
of samples per symbol grows with the square root of the SNR,
the capacity pre-log is at least 3/4.
The result in \cite{Ghozlan2014ISIT} is extended in \cite{barletta2015upper} to consider all scaling of the oversampling coefficient of the form $P^{\al}$.
Further simulations  to compute lower bounds on the information rates achieved by the
multi-sample receiver have been recently shown in  \cite{ghozlan2017models, nedelcu2017}.

\subsubsection*{Contribution}
In this paper, we investigate the capacity of the point-to-point channel corrupted by Wiener phase noise and additive white
Gaussian noise  with an   integrate-and-dump   multi-sample   receiver, which we refer to as oversampled Wiener phase noise (OWPN) channel.
Our main contributions are described as follows:

\smallskip
\noindent
$\bullet$ \textbf{Sec. \ref{Sec:outer}-- Capacity outer bound:}
Using the I-MMSE relation \cite{guo2005} and a lower bound on the minimum mean-square error (MMSE) estimate  expressed through a recursive equation \cite{Tichavsky1998},
we obtain a novel outer bound on the capacity of the OWPN channel.

\smallskip
\noindent
$\bullet$ \textbf{Sec. \ref{Sec:main}--Degrees of Freedom:}
The outer bound in Sec.  \ref{Sec:outer} is shown to be tight at high SNR; more specifically  we derive the GDoF for the model in which the transmit power
$P$, and the oversampling factor $L$, grow large for $L=\lfloor P^{\al} \rfloor$.

\subsubsection*{Paper Organization}

The channel model is presented in Sec.~\ref{sec:channel model} while the known results in the literature are presented in Sec.~\ref{sec:Known Results}.
Outer bounds are derived in Sec.~\ref{Sec:outer}, while the degrees of freedom analysis is shown in Sec.~\ref{Sec:main}.
Conclusions are drawn in Sec.~\ref{Sec:conclusion}.

\section{System Model}
\label{sec:channel model}
We consider the  OWPN, that is  the
point-to-point channel corrupted by Wiener phase noise and additive white
Gaussian noise  with
an   integrate-and-dump   multi-sample   receiver.
The main assumptions are as in \cite{ghozlan2013multi,barletta2015upper}, that is (i) the amplitude fading component, obtained by low-pass filtering the continuous time Wiener process,
is neglected by setting it to one, and (ii) the complex envelope of the transmitted waveform is constant in each symbol time interval.
%
%Following the formulation in \cite{ghozlan2013multi,barletta2015upper},
Under these assumptions, the channel output for this model is obtained as
\begin{equation}\label{eq:channel}
Y_{mL+l} = X_{m} e^{j \Theta_{mL+l}} + W_{mL+l},
\end{equation}
for $m \in [1,\ldots,M]$ and  $l \in [0,\ldots, L-1]$ and
where $W_{j} \sim {\cal CN}(0,2)$ is the additive noise, and $L\in \Nbb$ is the oversampling factor, that is equal to the reciprocal of the sampling time by assuming that the symbol time is unitary. The Wiener phase noise process $\{\Theta_j\}$ is defined as
\eas{
& \Theta_{L-1}\sim \Ucal([0,2 \pi]) \\
& \Theta_{k+1} = \Theta_{k} + N_k \\
& N_k \sim {\cal N}(0, \sgs L^{-1}),\quad k \in [L...\infty),
}{\label{eq:wiener}}
where $\Ucal(I)$ indicates the uniform distribution over the set $I$.
The channel input is subject to the power constraint
\ea{
\Ebb[|X_m|^2] \leq L^{-1}P,
\label{eq:power const}
}
which correspond to an average power constraint of $P$ for the transmitted waveform. As a consequence, the SNR is equal to $P/2$.
Define $\Yv_m= Y_{mL}^{(m+1)L-1}=[Y_{mL}, \ldots, Y_{(m+1)L-1}]$
and  $\Wv_m= W_{mL}^{(m+1)L-1}$: with this notation, the capacity of the channel in  \eqref{eq:channel} can be expressed as%
\ea{
\Ccal(P,\sgs,L)=\lim_{M \goes \infty} \frac{1}{M} \sup  I(\Yv^M;X^M)
\label{eq:capacity def}
}
where the supremum is over all the distributions of $X^M$ such that the power constraint $P$ is satisfied and for an oversampling factor equal to $L$.\footnote{In the following, we indicate the dependency of $\Ccal$ to $[P,\sgs,L]$ only when necessary.}

We also consider the high-SNR asymptotics of the expression in  \eqref{eq:capacity def} which are described by the GDoF, defined as
\ea{
D(\al)=\lim_{P \goes \infty} \f{\Ccal(P,\sgs,\lfloor P^{\al} \rfloor)} {\log(P)},
\label{eq:gdof def}
}
that is, the capacity pre-log factor when  $P$ grows to infinity while $L=\lfloor P^{\al} \rfloor$.

\smallskip

Note that, in the above formulation, the additive noise variance is not affected by the oversampling factor, while the transmit power of a sample is.
The detailed derivation of the  discrete-time model in~\eqref{eq:channel} from the continuous-time one is presented in~\cite{ghozlan2017models}.
Since $P$ and SNR are directly related, the degrees of freedom formulation in \eqref{eq:gdof def} correctly captures the asymptotic behaviour of capacity at high SNR.

\section{Known Results}
\label{sec:Known Results}

The OWPN encompasses the classic discrete-time Wiener phase noise (WPN) channel as the special case in which \mbox{$L=1$}.
We have recently derived the capacity of the WPN channel to within a small additive gap.

\begin{thm}{\bf Capacity bounds on the WPN channel \cite[Th. V.1]{barletta2017capacity}.}
\label{th:Capacity to within a constant gap}
The capacity of the WPN channel is upper-bounded as
\begin{align}
\Ccal &\leq \f 1 2 \log(1+P/2)\nonumber\\&+ \lcb  \p{
\f 1 2 \log(4\pi e)+ 2 \f {e^{-\frac{2\pi}{e}}}{1-e^{-\frac{2\pi}{e}}}\log(e) &    \sgs>\f {2 \pi}{e}  \\
%
%\f 12 \log(1 + P)-h([\sgs]_{2\pi})
 \f 12 \log \lb  \f 2 {   \sgs}  \rb +\log(2\pi) +\log^2(e)
&   P^{-1} \leq \sgs \leq \f {2 \pi}{e} \\
\f 12 \log(1+P/2)    &  P^{-1}>\sgs,
} \rnone
\label{eq:outer}
\end{align}
and the exact capacity is to within $ \Gcal \ \bpcu$ from the outer bound in \eqref{eq:outer}, where
\ea{
\Gcal \leq \lcb\p{
4 &    \sgs>\f {2 \pi}{e},\\
7.36 &   P^{-1} \leq \sgs \leq \f {2 \pi}{e}, \\
1.8 &  P^{-1}>\sgs.
}
\rnone
\label{eq:regimes phase noise}
}
\end{thm}

The result in Th. \ref{th:Capacity to within a constant gap}  is interesting at it shows that the capacity of the WPN channel can be sub-divided in three regimes:
(i)~for large  values of the
frequency noise variance~$\sigma^2$, the channel behaves similarly to a channel with circularly uniform iid phase noise;
(ii) when the
frequency noise variance is small,
the effect of the additive noise dominates over that of the phase  noise, while
(iii) for intermediate values of the
frequency noise variance, the transmission rate over the phase modulation channel
 has to be reduced due to the presence of phase noise.

A lower bound on the GDoF of the OWPN channel for $\al=1/2$ is obtained in \cite{ghozlan2014phase} and is later extended  in~\cite{barletta2015upper} to yield an  inner bound to the 
GDoF region.\footnote{The original result is derived for $\al \in [0,1]$ but can be easily extended to the case of $a \in \Rbb^+$.}

\begin{thm}{\bf GDoF lower bound \cite{ghozlan2014phase,barletta2015upper}.}
\label{th:lower gdof}
The function $D(\al)$ in \eqref{eq:gdof def} can be lower-bounded as
\ea{
D(\al) \geq \lcb \p{
\frac{1+\alpha}{2}       & 0 \leq \al < \f 12 \\
3/4        &  \al\ge \f 1 2,
}\rnone
\label{eq:lower gdof}
}
\end{thm}
The inner bound in Th. \ref{th:lower gdof} is obtained by letting the  channel input
 have uniformly distributed phase in $[0,2\pi]$ while the amplitude has a  shifted exponential distribution.
 At the receiver, the statistic used for detecting $|X_k|$ is $||\Y_k||$, and the one used for detecting $\phase{X_k}$
is $\angle\lb Y_{kL} \lb Y_{kL-1} e^{ -j\angle{X_{k-1}}}\rb^\star\rb$.

An outer bound on the capacity of the OWPN channel is derived in  \cite{barletta2015upper} which, together with Th. \ref{th:lower gdof}, yields the exact
GDoF expression for $\al \in [0,1/2)$.
\begin{thm}{\bf OWPN channel outer bound \cite{barletta2015upper}.}
\label{th: old OWPN channel outer bound}
The capacity of the OWPN channel is upper-bounded as
\ea{
\Ccal \leq \f 1 2 \log \lb 1 + \f P 2 \rb +\f 12 \log \lb \f {2 \pi}{e \sgs L^{-1}}  \rb+ \Ocal(1).
\label{eq: old OWPN channel outer bound}
}
\end{thm}

Combining the results in  \cite{ghozlan2014phase} and \cite{barletta2015upper}, for $\al \in[0,1/2]$ we obtain that
\ea{ \label{eq:dgof_known}
{\rm if} \  \al \in[0,1/2], \quad {\rm then} \ D(\al)=\f {1+\al} 2.
}
In the next section, we derive an outer bound tighter than that in Th. \ref{th: old OWPN channel outer bound} which yields the GDoF region for any $\al \in \Rbb^+$.

\section{Outer Bound}
\label{Sec:outer}
We begin by deriving an outer bound on the capacity which improves over the result in Th.~\ref{th: old OWPN channel outer bound}. Specifically,
we provide a better estimate of the transmission rate that can be attained through phase modulation of the channel input.
\begin{thm}{\bf Capacity Outer bound.}
\label{th:Capacity Outer bound}
The capacity of the OWPN channel is upper-bounded as
\ea{
\Ccal &  \leq \f 1 2 \log \lb 1 + \f P 2 \rb +\log \lb 2\pi \rb
\label{eq:Capacity Outer bound}\\
& \quad \quad +\f 12 \log\left(L^{-1} P \left(\sqrt{1+4 \frac{1}{\sgs L^{-2} P}}-1\right) \right).
\nonumber
}
\end{thm}
\begin{IEEEproof}
Let us begin by upper bounding the information rate in  \eqref{eq:capacity def} as
\ea{
& \mi{X_1^M}{\Y_1^M}  \nonumber \\
&=\sum_{k=1}^{M} \micnd{X_1^M}{\Y_k}{\Y_1^{k-1}} \nonumber\\
&\le \sum_{k=1}^{M} \micnd{X_1^M, \Theta_{kL-1}}{\Y_k}{\Y_1^{k-1}} \nonumber\\
&= \sum_{k=1}^{M} \micnd{X_k}{\Y_k}{\Theta_{kL-1}} + \micnd{\Theta_{kL-1}}{\Y_k}{\Y_{1}^{k-1}}   \label{eq:upper1}
%	\nonumber\\
}
where \eqref{eq:upper1} follows from the Markov chain $\Yv_k\markov X_k,\Theta_{kL-1}\markov\Y_{1}^{k-1}$.
Since the noise is circularly symmetric, a circularly distributed input is capacity achieving: accordingly we have
\begin{align}
& \micnd{\Theta_{kL-1}}{\Y_k}{\Y_{1}^{k-1}}  \nonumber \\
&\le \I \lb  \Theta_{kL-1}; \left\{\Theta_{kL-1}\oplus \phase{X_k}\oplus\sum_{i=0}^{\ell-1} N_{kL+i-1}\right\}_{\ell=1}^L \rnone \nonumber \\
& \quad \quad \lnone \lnone  |X_k|, \Wv_k  \rabs \  \Y_{1}^{k-1} \rb \nonumber \\
&=\micnd{\Theta_{kL-1}}{ \Theta_{kL-1}\oplus \phase{X_k}\oplus N_{kL-1}, |X_k|, \Wv_k  }{\Y_{1}^{k-1}} \nonumber\\
&=0 \label{eq:mithetaY1}
\end{align}
where \eqref{eq:mithetaY1} follows from the fact that the input is circularly symmetric, so that $|X_k| \perp  \phase{X_k}$, and independent of the phase $\Theta_{kL-1}$.
Similarly to \cite[Eq.~(19)]{barletta2015upper}, we note that the term $\micnd{X_k}{\Y_k}{\Theta_{kL-1}}$
 can be divided into two contributions:  one from the channel input amplitude and the other from channel input phase.
 In fact, using \eqref{eq:mithetaY1}, we can write
\ea{
 & \frac{1}{M}\mi{X_1^M}{\Y_1^M}
 \leq \frac{1}{M}\sum_{k=1}^{M} \micnd{X_k}{\Y_k}{\Theta_{kL-1}}\nonumber\\
 &=I(X_1,\Yv_1|\Theta_{L-1}) \nonumber\\
 &=\micnd{|X_1|}{\Y_1}{\Theta_{L-1}}+\micnd{\phase{X_1}}{\Y_1}{\Theta_{L-1},|X_1|} \label{eq:polar}
}
where the first equality follows from stationarity of the processes, and the last step by polar decomposition of $X_1$.
In the following, we refer to $\micnd{|X_1|}{\Y_1}{\Theta_{L-1}}$ as the \emph{rate of the amplitude channel} and $\micnd{\phase{X_1}}{\Y_1}{\Theta_{L-1},|X_1|}$ as the \emph{rate of the phase channel}.
Analogously to~\cite[Eq.~(20)]{barletta2015upper}, the rate of the amplitude channel rate can be bounded as
\eas{
&  \micnd{|X_1|}{\Y_1}{\Theta_{L-1}} \nonumber\\
& \leq \micnd{|X_1|}{\Y_1,\Theta_L^{2L-1}}{\Theta_{L-1}} \nonumber\\
&=\mi{|X_1|}{ \labs \sqrt{L} X_1 + \f 1 {\sqrt{L}} \sum_{\ell=0}^{L-1} W_{L+\ell} \rabs }
 \label{eq:abs}   \\
&\le \f 12 \log(2\pi e(  P+2 )),
\label{eq:last no over}
}{\label{eq:amplitude subchannel}}
where \eqref{eq:last no over} follows from the result in~\cite[Thm. IV.1]{barletta2017capacity} which bounds the entropy of a non-central chi-square random variable.

The rate in the phase modulation channel can be written as
\eas{
	& \micnd{\phase{X_1}}{\Y_1}{\Theta_{L-1},|X_1|} \nonumber \\
	& = \entcnd{\phase{X_1}}{\Theta_{L-1},|X_1|} - \entcnd{\phase{X_1}}{\Theta_{L-1},|X_1|, \Y_1} \\
	&= \log(2\pi) -\entcnd{\phase{X_1}}{\Theta_{L-1},|X_1|, \Y_1},
\label{eq:phase_subchannel}
}
where \eqref{eq:phase_subchannel} follows from the fact that $\phase{X_k}\sim {\cal U}([0,2\pi))$.
The entropy term in \eqref{eq:phase_subchannel} can be rewritten as
\eas{
& -\entcnd{\phase{X_1}}{\Theta_{L-1},|X_1|, \Y_1} \nonumber \\
& = -\entcnd{\phase{X_1}}{|X_1|, \Y_1 e^{- j \Theta_{L-1} }} \label{eq:pp0} \\
& = \h \lb \Theta_{L-1} | \  |X_1|, \rnone
\label{eq:pp1} \\
& \quad \quad \quad \lnone \left\{|X_1| \exp\left(j\left(\Theta_{L-1}+\sum_{i=L-1}^{k-1} N_i\right)\right)+W_k \right\}_{k=L}^{2L-1} \rb \nonumber \\
&=- \entcnd{\Theta_{L-1}}{|X_1|, \widetilde{Y}_{L}^{2L-1}}
\label{eq:pp2}\\
& \leq - \entcnd{\Theta_{L-1}}{|X_1|, \widetilde{Y}_{L}^\infty},
\label{eq:pp3}
}{\label{eq:pp}}
where $\widetilde{Y}_k = |X_1| e^{j \Theta_k}+W_k$. Here,  \eqref{eq:pp0} follows from the fact that $\Theta_{L-1} \sim \Ucal([0,2 \pi))$ and independent of all other variables.
In \eqref{eq:pp1} we let $\phase{X_1}=\Theta_{L-1}$: this substitution is to stress the fact that  $\phase{X_1}$ is independent of all other variables and is thus
statistically equivalent to $\Theta_{L-1}$.
In other words, the entropy of the phase of $X_1$ given the knowledge  of $\Theta_{L-1}$ and $(Y_{L}^{2L-1},|X_1|)$ is  equivalent to the entropy of
$\Theta_{L-1}$ given $(\widetilde{Y}_L^{2L-1},|X_1|)$.
Finally,  \eqref{eq:pp3} follows from the ``conditioning reduces entropy'' property.

From the I-MMSE relationship~\cite[Eq.~(182)]{guo2005}, we have
\ea{
& \entcnd{\Theta_{L-1}}{|X_1|, \widetilde{Y}_{L}^\infty} = \f 12\log(2\pi e \variance{\Theta_{L-1}})\nonumber\\
&  \quad - \f 12 \int_{0}^{\infty} \left[\frac{\variance{\Theta_{L-1}}}{1+\rho \variance{\Theta_{L-1}}} \rnone \nonumber \\
&  \quad  \quad \quad \lnone - \text{mmse}(\Theta_{L-1}| \sqrt{\rho} \Theta_{L-1} + Z, |X_1|, \widetilde{Y}_{L}^\infty)\right] d\rho, \label{eq:I-MMSE}
}
where $Z\sim {\cal N}(0,1)$ and independent of any other quantity, and
\begin{align}
\text{mmse}(S|K) \triangleq \expect{(S-\expcnd{S}{K})^2}.
\end{align}
The crucial step in bounding the entropy term in \eqref{eq:pp3} using the relationship in \eqref{eq:I-MMSE} is in obtaining a tight lower bound to the MMSE.
To obtain such lower bound we rely on the result in \cite[Prop.~1]{Tichavsky1998}.
To this  end, let us rewrite $U=\sqrt{\rho} \Theta_k + Z$ and
\begin{align}
& \text{mmse}(\Theta_{L-1}| \sqrt{\rho} \Theta_{L-1} + Z, |X_1|, \widetilde{Y}_{L}^\infty) \\
&= \lim_{k\rightarrow\infty} \text{mmse}(\Theta_k | U, |X_1|, \widetilde{Y}_1^{k-1}) \nonumber\\
&\ge \lim_{k\rightarrow\infty} J_{k}^{-1},
\end{align}
where we used the time-reversibility of the Wiener process and the Bayesian Cramer-Rao inequality. Here $J_k$ is  defined as
the entry in position $(k,k)$ of the information matrix associated with the joint distribution of $[ \Theta_1^k, \widetilde{Y}_1^{k-1},U]$ given $|X_1|$.
According to~\cite[Prop.~1]{Tichavsky1998}, the value $J_k$ can be computed recursively as
\begin{align}\label{eq:Jk}
J_k = D_{k-1}^{(22)} - \frac{(D_{k-1}^{(12)})^2}{J_{k-1}+D_{k-1}^{(11)}},
\end{align}
for
\eas{
D_n^{(11)} & =  \frac{L}{\sgs }, \quad n=1,\ldots,k-1 \label{eq:D11} \\
D_n^{(12)} & = -\frac{L}{\sgs}, \quad n = 1,\ldots,k-1 \\
D_{k-1}^{(22)} & =  \frac{L}{\sgs}+\rho, \\
D_{n}^{(22)} &= \frac{L}{\sgs}+\expect{|X_1|^2} \nonumber\\
&\le \frac{L}{\sgs}+L^{-1} P, \qquad n=1,\ldots,k-2, \label{eq:D22}
}{\label{eq:D}}
where \eqref{eq:D22} follows from  the average power constraint. Using~\eqref{eq:D} into~\eqref{eq:Jk} we can easily find  a lower bound on $J_\infty$.
Using this bound in~\eqref{eq:I-MMSE} and then in~\eqref{eq:phase_subchannel}, an upper bound on the rate of the phase channel is obtained as
\ea{
& \micnd{\phase{X_1}}{\Y_1}{\Theta_{L-1},|X_1|}
	\le\f 12\log \lb \frac{2\pi}{e}\rb   \label{eq:final_phase}  \\
& \quad + \f 12 \log\left(\frac{L^{-1} P}{2} \left(\sqrt{1+4 \frac{1}{\sgs L^{-2}P}}-1\right) \right).
\nonumber
}
Combining \eqref{eq:last no over} and \eqref{eq:final_phase} we obtain an outer bound on capacity as in~\eqref{eq:Capacity Outer bound}.
\end{IEEEproof}

\section{Degrees of Freedom Analysis}
\label{Sec:main}
The outer bound in Th. \ref{th:Capacity Outer bound}, together with the results in Th.~\ref{th: old OWPN channel outer bound} and Th.~\ref{th:lower gdof} yields the GDoF
of the OWPN channel.
\begin{lem}
\label{lem:dgof}
The GDoF of the OWPN channel is obtained as
\ea{ \label{eq:dgof}
D(\al)=\lcb\p{
\f {1+\al} 2  & 0\le \al \le \f 1 2,\\
\f 3 4  &  \al > \f 1 2.
}
\rnone
}
\end{lem}
\begin{IEEEproof}
As shown in \eqref{eq:dgof_known}, the GDoF in known for $\al \in [0,1/2]$. For $\al>1/2$ consider the outer bound in Th. \ref{th:Capacity Outer bound} for
$L = \lfloor P^\alpha \rfloor$:
for the rate in the amplitude modulation channel in \eqref{eq:amplitude subchannel}, we have
\ea{
\label{eq:dgof_amplitude}
& \lim_{P\rightarrow\infty} \frac{ \micnd{|X_1|}{\Y_1}{\Theta_{L-1}}}{\log(P)}\\
&  \quad \quad  \leq  \lim_{P\rightarrow\infty}   \f 12 \f {\log(2\pi e(  P+2 ))} {\log P}=\frac{1}{2}, \nonumber
}
which holds for any $\al \in \Rbb^+$.
For the rate in the phase modulation channel in \eqref{eq:final_phase} we have
\ea{\label{eq:dgof_phase}
& \lim_{P\rightarrow\infty} \frac{\micnd{\phase{X_1}}{\Y_1}{\Theta_{L-1},|X_1|}}{\log(P)}  \\
&  \quad \quad    \leq \lim_{P\rightarrow\infty}  \f 12 \f {\log\left(\frac{L^{-1} P}{2} \left(\sqrt{1+4 \frac{1}{\sgs L^{-2}P}}-1\right) \right)}{\log P} = \f 14, \nonumber
}
which follows from the fact that, for $\al> 1/2$, we have that $L^{-2} P=P^{1-2\alpha} \rightarrow 0$ as $P\rightarrow\infty$.

Combining \eqref{eq:dgof_amplitude} and \eqref{eq:dgof_phase} we obtain  the outer bound on $D(\al)$ with matches the inner bound in \eqref{eq:lower gdof}
for the regime $\al \in [1/2,\infty)$.
\end{IEEEproof}

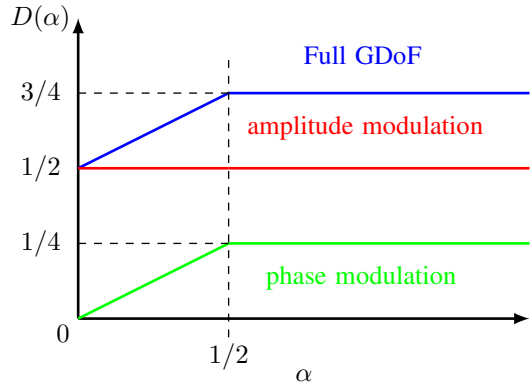
\begin{figure}
\centering
\begin{tikzpicture}[node distance=2cm,auto,>=latex]
  \draw[->,line width=1pt] (0,0) -- (0,4);
  \node at (-.5,4) (ly) {$D(\al)$} ;
  \draw[->,line width=1pt] (0,0) -- (6,0);
  \node at (3,-.75) (lx) {$\al$} ;
  \draw[line width=1pt, color=blue] (0,4/2) -- (4/2,6/2);
  \draw[-,line width=1pt,color=blue] (4/2,6/2) -- node[above, xshift = 0.25cm, yshift = 0.25cm, text width  = 2.5 cm,color=blue] {Full GDoF} (6,6/2);
  \draw[line width=1pt, color=green] (0,4/2-2) -- (4/2,6/2-2);
  \draw[-,line width=1pt,color=green] (4/2,6/2-2) -- node[above, xshift = 0.5cm, yshift = -0.75cm, text width  = 4 cm,color=green] {phase modulation } (6,6/2-2);
  \draw[line width=1pt, color=red] (0,2) -- (4/2,2);
  \draw[-,line width=1pt,color=red] (4/2,2) -- node[above, xshift = 0.75cm, yshift = 0.25cm, text width  = 5 cm,color=red] {amplitude modulation } (6,2);
  \draw[-,line width=.5 pt,color=black, dashed] (0,6/2-2) --  (4/2,6/2-2);
  \draw[-,line width=.5 pt,color=black, dashed] (0,6/2) --  (4/2,6/2);
  \draw[-,line width=.5 pt,color=black, dashed] (2,-0.25) --  (4/2,7/2);
      \node at (2,-.5) (lx) {$1/2$};
  \node at (-0.5,4/2) (lx) {$1/2$};
  \node at (-0.5,6/2-2) (lx) {$1/4$};
  \node at (-0.5,6/2) (lx) {$3/4$};
  \node at (-0.2,-0.2) (lx) {$0$};
\end{tikzpicture}
\caption{
The degrees of freedom of the OWPN channel in Lem.~\ref{lem:dgof}.
}
\label{fig:gdof}
\end{figure}

The result in Lem. \ref{lem:dgof} is schematically represented in Fig.~\ref{fig:gdof}:
the GDoF from amplitude modulation are equal to $1/2$ for all $\al$, while the GDoF from phase modulation are equal to $(1+\al)/2$ for $\al \in [0,1/2]$ and equal
 to $1/4$ for $\al>1/2$.
Note that, when $\al \goes 0$, we obtain the model with $L=1$ in Th. \ref{th:Capacity to within a constant gap} which has pre-log equal to $1/2$.

\subsubsection*{Discussion}
The analysis of the GDoF in Lem. \ref{lem:dgof} suggests that there is a fundamental tension between the AWGN and the multiplicative WPN, and improving the resolution of the receive
 filter beyond $L^{-1}=1/\sqrt{P}$ does not improve the capacity pre-log at large~$P$. 
From a high level perspective, the parameter  $\sigma^2$ is related to the quality of the local oscillators available at the user: in this sense, then, the result in Lem. \ref{lem:dgof} shows that, regardless of the value $\sgs$,
the fundamental tension will eventually reduce the available DoF for a suitably large $P$.

From the I-MMSE bound in \eqref{eq:I-MMSE} and \eqref{eq:Jk} used in the proof of Th.~\ref{th:Capacity Outer bound}, it is apparent that the tension between the AWGN and the WPN 
is related to the difficulty of predicting a new sample of a Wiener process when corrupted by AWGN.
The following questions naturally arise: is the limitation of the available GDoF an artifact of the assumptions used to derive the model in~\eqref{eq:channel} or is it an inherent limitation 
of the physical system?
Further, the model in~\eqref{eq:channel} neglects the effect of amplitude fading  for the sake of simplicity. 
For the model encompassing both phase and amplitude fading, one wonders whether it is possible to attain higher DGoF.
The model in \eqref{eq:channel} is obtained by employing a waveform that allocates the power uniformly over time.
One then naturally wonders whether it is possible to attain higher performance employing  a waveform  that does not allocate energy uniformly in time.
These interesting open questions are to be addressed in future works.

\section{Conclusion}
\label{Sec:conclusion}
We have derived an outer bound on the capacity of discrete-
time Wiener phase noise channels with multi-sample receivers. 
In this model, the input of a point-to-point channel is corrupted by both additive  noise and multiplicative phase noise: 
the additive noise is a white Gaussian process while the phase noise is a Wiener process.
For each symbol in input, the channel produces $L$ outputs corresponding to the output of an integrate-and-dump multi-sample receiver with oversampling factor $L$.
A novel outer bound is derived  using the I-MMSE relationship and a recursive expression of the minimum mean-square error though the Fisher information matrix.
This novel outer bound is used to derive the generalized degrees of freedom for the scenario in which the oversampling factor grows with the transmit 
power $P$ as $L=\lfloor P^{\al}\rfloor$.
This latter result shows that there exists a fundamental tension between the AWGN and the  WPN that limits the available GDoF at $1/4$ for phase modulated signals,
 regardless of the power of the phase noise.
 The degrees of freedom analysis of models encompassing both multiplicative phase noise and multiplicative amplitude noise remains an interesting open question.

\bibliographystyle{IEEEtran}
\bibliography{steBib1}

\end{document}